\documentclass[number,sort&compress]{elsarticle}

\usepackage{graphicx}
\usepackage{amsmath}
\usepackage{amssymb}

\begin{document}

\begin{frontmatter}
\title{When atomic-scale resolution is not enough:\\ Spatial effects in {\em in situ} model catalyst studies}

\author[tum]{Sebastian Matera\corref{cor1}}
\ead{sebastian.matera@ch.tum.de}
\author[tum]{Karsten Reuter}

\address[tum]{Department Chemie, Technische Universit{\"a}t M{\"u}nchen, Lichtenbergstr. 4, D-85747 Garching, Germany}
\cortext[cor1]{Corresponding author, Fax: +49 89 289 13622 }

\begin{abstract}
We investigate transport effects in {\em in situ} studies of defined model catalysts using a multi-scale modeling approach integrating first-principles kinetic Monte Carlo simulations into a fluid dynamical treatment. We specifically address two isothermal flow setups: i) a channel flow with the gas-stream approaching the single crystal from the side, as is representative for reactor scanning tunneling microscopy experiments; and ii) a stagnation flow with perpendicular impingement. Using the CO oxidation at RuO$_2$(110) as showcase we obtain substantial variations in the gas-phase pressures between the inlet and the catalyst surface. In the channel geometry the mass transfer limitations lead furthermore to pronounced lateral changes in surface composition across the catalyst surface. This prevents the aspired direct relation between activity and catalyst structure. For the stagnation flow the lateral variations are restricted to the edges of the catalyst. This allows to access the desired structure-activity relation using a simple model.
\end{abstract}


\end{frontmatter}

\section{Introduction}

The Surface Science approach focusing on model single-crystal catalysts under ultra-high vacuum (UHV) conditions has provided a firm basis for our atomic-scale understanding of heterogeneous catalytic processes. Notwithstanding, aside from the obvious materials gap also potential pressure gap effects are a continuous source of concern when aiming to transfer these insights to technological conditions. In order to scrutinize this point a range of {\em in situ} approaches to study model catalysts has recently been pushed forward \cite{stierle2007}. Next to controlled kinetic measurements at near- and above ambient pressure conditions, these are notably local {\em in situ} microscopies and spectroscopies like surface X-ray diffraction (SXRD) \cite{ferrer2007}, X-ray photoelectron spectroscopy (XPS) \cite{bluhm2007} or scanning tunneling microscopy (STM) \cite{frenken2007}. As these techniques often impose significant constraints on the design of the reactor chamber, one immediate "pressure gap" effect that does not originate from the actual surface chemistry are  potential heat and mass transfer limitations in the fluid above the model catalyst.

In accordance with recent experimental findings \cite{gao2009,vanrijn2011,chen2010}, we have emphasized the crucial role of such limitations in this emerging field before \cite{matera2009,matera2010} -- in particular for unselective and therefore high turnover processes like the CO oxidation that are preferentially used as allegedly simple test reactions. Integrating a first-principles microkinetic model of CO oxidation at RuO$_2$(110) \cite{reuter2004,reuter2006} into fluid dynamical simulations, we illustrated how critically heat and mass transfer limitations can mask the intrinsic activity for an idealized stagnation flow reactor. In this geometry, the gas stream impinges perpendicularly onto the flat-faced single-crystal surface. This is a desirable flow scenario for controlled measurements, as it ensures that (apart from edge effects) the dominant part of the active surface sees at least the same gas-phase composition -- albeit possibly not the nominal one due to heat and mass transport limitations. 

Nevertheless, this idealized flow geometry is rarely accomplished in real experimental setups, where e.g. pumps in case of {\em in-situ} XPS \cite{yamamoto2008} or the tip in case of reactor STM \cite{rasmussen1998} block the area above the catalyst. Particularly the latter reactor STM geometry is much better approximated as a lateral channel flow, in which the gas streams over the active surface from the side. Moreover, this flow scenario represents in some respect also an opposite extreme to the previously considered stagnation flow. Comparison of the insights obtained from these two extremes allows therefore to some extent a critical discussion of {\em in situ} setups in general. With this motivation we here advance our first-principles multi-scale methodology to two-dimensional (2D) flow geometries in order to investigate the channel flow and the influence of edge effects for the stagnation flow. As before we stick to the established first-principles microkinetic model of CO oxidation over RuO$_2$(110) \cite{reuter2004,reuter2006} as a suitable showcase. 

As for the stagnation flow before, we also obtain for the channel flow significant deviations of the gas-phase composition over the model catalyst away from the commonly accessible compositions at the inlet or at the outlet. This time, however, these deviations extend also to the lateral position at the active surface, i.e. different areas of the single-crystal catalyst see notably different environments. In turn, integrally measured overall activities only represent an average over the different local activities corresponding to the local surface composition and inherent activity. Simultaneously, local {\em in-situ} spectroscopies may yield radically different insights, depending on where they are exactly positioned over the crystal. This raises serious concerns about this particular reactor geometry and other {\em in situ} setups with flow profiles somewhere between this and the stagnation flow regime. In contrast, in the spatially resolved stagnation flow such lateral variations are restricted to the edges of the catalyst, while its dominant central part sees the same gas phase conditions. This enables a simple analysis of real-life experiments by means of a simple model for the idealized stagnation flow. While this geometry may be unsuitable for particular {\em in-situ} experiments, it at least allows to obtain the intrinsic reactivity as a function of the reaction conditions at the surface, which is a formidable task for less controlled flow profiles like the channel flow.

These findings highlight the difficulty of comparing data obtained in {\em in situ} setups with different flow geometries. In all cases, serious mass transfer effects seem to complicate the quest for a molecular-level understanding of catalytic processes in technological environments. Next to further improving the resolving power of the various {\em in-situ} techniques, further progress along this route will therefore critically depend on overcoming such flow limitations in experimental reactor setups. Under all circumstances it seems that {\em in-situ} experiments must be complemented by a detailed analysis of the gas phase transport in order to extract the relevant information and to make experiments comparable.

\section{Theory}

\subsection{Reactor geometries and velocity flow profiles}

\begin{figure}
\includegraphics[width=\linewidth]{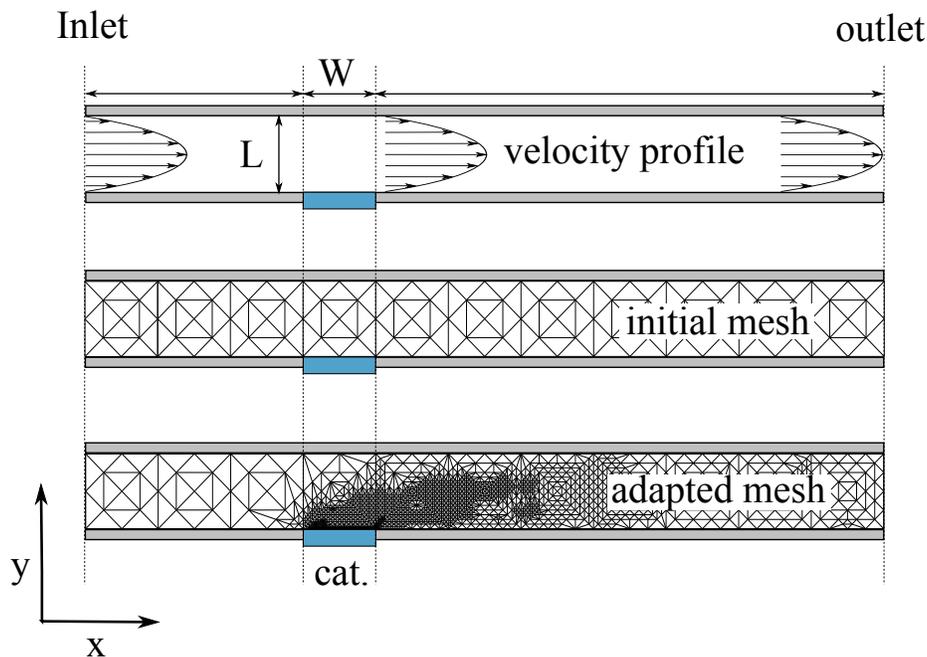}
\caption{(Color online) 2D channel flow geometry with a channel width $L$ and with the flat-faced model catalyst of width $W$ (depicted in blue) embedded in the bottom reactor wall. The gas streams from the inlet at the left along the $x$-direction to the outlet at the right. Additionally shown is the Hagen-Poiseuille parabolic velocity flow field, as well as the initial (middle panel) and final (lower panel) grid for the adaptive FEM simulation.}
\label{fig1}
\end{figure}

For the channel flow we consider the 2D geometry shown in Fig. \ref{fig1}, where the model catalyst is embedded into the lower planar reactor wall. The gas streams from the inlet at the left along the $x$-direction over the catalyst of width $W$ to the outlet to the right. In the $y$-direction the channel has a width $L$ and we neglect any influence of the third spatial dimension. Despite the heat released by the exothermic surface reactions an ideal heat coupling of the catalyst to the outside system maintains the nominal temperature $T$ throughout. For this isothermal limit the continuum mechanical description of the mass transport through the channel centers on solving the equations of motion for the $(x,y)$-fields of density $\rho$, velocity ${\bf v}$ and mass fractions $Y_\alpha$ (with $\alpha$ representing the involved species (O$_2$, CO, CO$_2$)). In the present context this description is simplified by the absence of relevant gravitational effects, by low flow velocities leading to laminar flows, and the possibility to work in the Low-Mach-Number-Approximation (LMA) \cite{majda1985}. Furthermore we want to postulate the common non-slip boundary conditions, i.e. the velocity components tangential to the reactor walls are zero there. Assuming an initially well mixed gas and without significant gas-phase chemical conversions in low-temperature CO oxidation the velocity field upstream of the catalyst will then exhibit the typical Hagen-Poiseuille parabolic profile depicted in Fig. \ref{fig1},
\begin{eqnarray}
\label{velocityAnalytical}
v_x &=& \left(-4\left( \frac{y}{L}\right)^2 \;+ \; 4\left( \frac{y}{L}\right)\right) u_{\rm max} \\ \nonumber
v_y &=& 0 \quad ,
\end{eqnarray}
with the maximal velocity $u_{\rm max}$ reached in the middle of the channel. For the present purpose we will keep this velocity profile also in the rest of the channel, i.e. also directly above the active catalyst \cite{maestri2008}. This essentially corresponds to assuming a constant viscosity and 
\begin{equation}
{\bf v} \cdot\nabla \rho \;\approxeq\; 0 \quad .
\label{approxMassBalance}
\end{equation}
The prior assumption is justified by the similar molecular weights and viscosities of the involved species O$_2$, CO, and CO$_2$. Under isothermal conditions catalytic conversions will then not change the overall viscosity much, as can e.g. directly be seen from common mixture-averaged viscosity theories \cite{bird1960}. As to Eq. (\ref{approxMassBalance}) we expect prominent gradients of $\rho$ only close to the catalyst surface. With an almost zero velocity there, the cross-coupling of velocity and composition fields induced by the term in Eq. (\ref{approxMassBalance}) will be negligible, as we will explicitly verify for the stagnation flow geometry below.

\begin{figure}
\includegraphics[width=\linewidth]{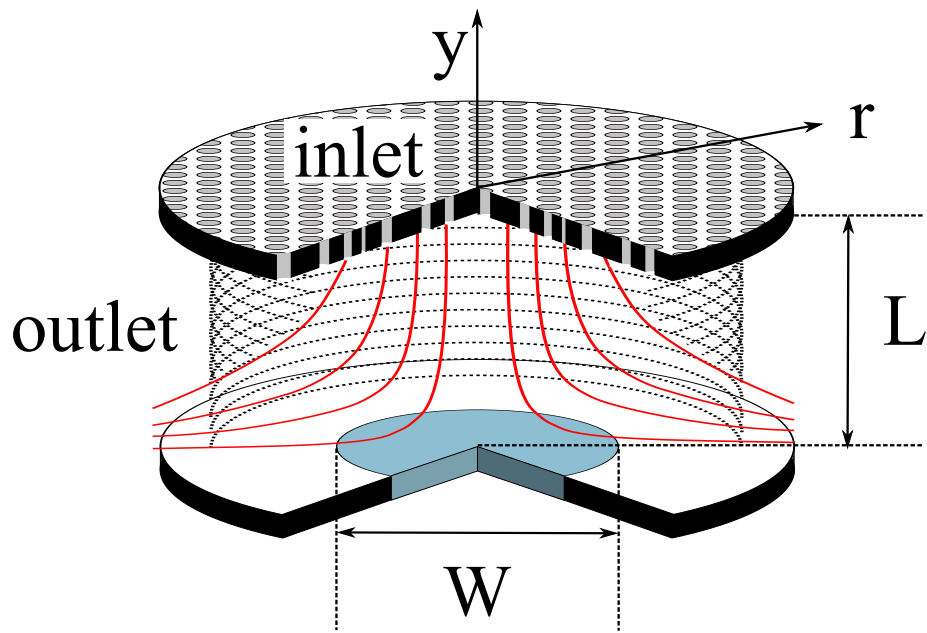}
\caption{(Color online) Stagnation flow geometry with the flat-faced model catalyst of width $W$ (depicted in blue) embedded in the disc at the bottom, and oriented parallel to the inlet located at a distance $L$ away. The gas streams with a uniform velocity at the inlet in the axial direction against the bottom wall. The gas stream splits at the symmetry axis $y$ transporting the gas to the cylindrical 
outlet. Additionally shown are typical stream lines illustrating the flow profile.}
\label{fig2}
\end{figure}

The stagnation flow geometry is depicted in Fig. \ref{fig2}. The gas mixture streams from a sieve-like inlet against a parallel disk located at a vertical distance $L$ from the inlet. The model catalyst with the diameter $W$ is smoothly integrated in the center of this disk. The red lines in Fig. \ref{fig2} represent a typical streamline profile for this stagnation flow. The rotational symmetry implies to work in radial coordinates with the axial coordinate $y$ and the radial coordinate $r$. For the following discussion the origin is placed at the center of the catalyst and $y$ points towards the inlet. The sieve-structure ensures that the velocity is uniform at the inlet with only the axial ($y$-) component $v_y(L,r)=-u_{\rm max}$ different from zero. As for the channel we consider a well-mixed gas phase at the inlet. As the outlet we consider an imaginary cylinder (as depicted by the thin dotted line in Fig. \ref{fig2}) between inlet and bottom disc with a radius larger than that of the catalyst. With the same arguments as for the channel we want to assume that the velocity field is not affected by the ongoing surface chemistry. Further assuming a sufficiently large diameter of the inlet and the bottom disk we can then approximate the velocity within the inlet, the surface and the outlet cylinder by the solution for an ideal nonreactive stagnation flow. Under these premises a similarity transform reduces the Navier-Stokes equations to a one-dimensional (1D) non-dimensionalized boundary value problem \cite{kee2003},
\begin{equation}
 \begin{array}{l}
 \vspace{0.2cm} \frac{d}{d \hat{y}} u = -2 V,\\ \vspace{0.2cm}
 \frac{d}{d \hat{y}} V = A,\\ \vspace{0.2cm}
\frac{d}{d \hat{y}} A = \frac{1}{Re}(uA + (V)^2 + \Lambda_r),\\ \vspace{0.2cm}
 \frac{d}{d \hat{y}} \Lambda_r=0,\\ \vspace{0.2cm}
 \text{with the boundary conditions}\\ \vspace{0.2cm}
 u(1)=-1,~ V(1)=0,~u(0)=0,~ V(0)=0,
 \end{array} \label{IncompStag}
\end{equation}
where $u$ is the scaled axial velocity, $V$ the scaled reduced radial velocity and $\Lambda_r$ is the so-called radial pressure curvature.
The Reynolds number $Re=\rho^{\rm inlet} u_{\rm max} L /\mu^{\rm inlet}$ is the only parameter describing the whole flow configuration, where $\rho^{\rm inlet}$ is the density at the inlet. The viscosity $\mu^{\rm inlet}$ is determined from the inlet concentrations and temperature, employing the same mixture-averaged formula as in Ref. \cite{matera2010}. The required velocity components at a point $(y,r)$ are then determined by rescaling: $\hat{y}=y/L$, $v_y(y,r)=u_{\rm inlet} u(\hat{y})$ and $v_r(y,r)=r V(\hat{y}) u_{\rm max}/L$. In practice we solve this boundary value problem (\ref{IncompStag}) for Reynolds numbers $Re \in [0.1,1000]$ using the collocation solver {\sc COLNEW} \cite{COLNEW} employing an error criterion of $10^{-8}$. We then interpolate its solution as a function of reduced axial coordinate $\hat{y}$ and $\log_{10} Re$ employing third order B-Splines \cite{BSplines} with 130 equidistant grid points in each direction.

For the width of the catalyst $W$ we choose 1\,cm in both cases. To make the results comparable for the two reactor geometries, the characteristic reactor size $L$ is set to same value of 1\,cm, and the same inlet velocity parameter $u_{\rm max}=20$\,cm/s is used in both cases. For the channel the inlet was conveniently placed 3\,cm upstream of the active catalyst. The outlet position needs to be sufficiently far downstream of the active catalyst to ensure consistency with the concentration equilibration assumed by the Neumann boundary condition Eq. (\ref{Neumann}) below. We found this well achieved by placing the outlet 7\,cm downstream of the catalyst' right edge. For the stagnation flow geometry the cylindrical outlet was taken to have a diameter of 10\,cm also ensuring consistency with the boundary conditions detailed below.

\subsection{Balance equations and boundary conditions}

With the velocity profiles given, what remains to be solved for both reactor geometries are the balance equations for the species' mass fraction $Y_\alpha$,
 \begin{eqnarray}
   \rho\partial_t {Y_\alpha} + \rho {\bf v}  \cdot \nabla Y_\alpha + \nabla \cdot {\bf j}_\alpha \;=\;  \tau_{\alpha}  \label{Species}\\      {\bf j}_\alpha=-\sum\limits_{\beta}  D_{\alpha \beta} \nabla Y_\beta  \label{Species2} \quad ,
 \end{eqnarray}
with the diffusive mass flux ${\bf j}_\alpha$. The diffusion coefficients $D_{\alpha \beta}$ are obtained through the Stefan-Maxwell equations for the mass fluxes ${\bf j}_\alpha$ as detailed in Ref. \cite{matera2010}. In the LMA the density $\rho$ is obtained from the (spatially) constant total reference pressure $p$ through the ideal gas law and is thus a mere function of the mass fractions. With no significant gas-phase chemical reactions for the considered low temperature CO oxidation the associated source term $\tau_{\alpha}$ is zero. Eqn. (\ref{Species}) and (\ref{Species2}) need to be complemented with appropriate boundary conditions at the reactor inlet and outlet, at the reactor walls and the catalyst's surface. At the inlets this is obviously the nominal chemical composition $Y_\alpha^{\rm inlet}$ of the reactant feed, together with the temperature $T$ and the total pressure $p$, that are constant throughout the system under the present isothermal and LMA approximations. If the outlet is placed sufficiently far away from the reactive zone we can furthermore assume that all concentration variations have equilibrated there, so that an appropriate Neumann boundary conditions is
\begin{equation}
{\bf n} \cdot \nabla Y_{\alpha}^{\rm outlet} \;=\;0 \quad ,
\label{Neumann}
\end{equation}
where $\bf{n}$ represents the normal vector of the boundary pointing into the reaction chamber.

The remaining crucial ingredients are the boundary conditions at the reactor walls and the catalyst's surface. Since there is no chemical conversion at the impermeable reactor walls, the normal mass fluxes are all zero there, leading by Eq. (\ref{Species2}) to the same Neumann boundary conditions as in Eq. (\ref{Neumann}). In contrast, at the catalytic surface there is a consumption of reactants and a source of products, in other words the normal mass fluxes are non-zero there. In the case of (quasi) stationary CO oxidation these obey the surface balance equation
\begin{equation}
  {\bf j}_\alpha^{\rm surf} \cdot {\bf n}\;=\;{\bf n}\cdot( -\sum\limits_{\beta}  D_{\alpha \beta} \nabla Y_\beta^{\rm surf})  \;=\; m_\alpha \nu_{\alpha} {\rm TOF} \quad ,
\label{surfaceSource}
\end{equation}
which provides a nonlinear Cauchy boundary condition. Here, $m_\alpha$ is the mass of a molecule of species 
$\alpha$, and $\nu_\alpha$ is its stoichiometric coefficient in the reaction ($\nu_{\rm O_2} = -1/2, \nu_{\rm CO} = -1, \nu_{\rm CO_2} = +1$ for the considered CO oxidation reaction). ${\rm TOF}$ is the turnover frequency of the reaction, i.e. the number of reactions taking place per time and surface area.
 
As in our previous work on the idealized stagnation flow we employ for this the established first-principles kinetic Monte Carlo (1p-kMC) model of CO oxidation at RuO$_2$(110) \cite{reuter2004,reuter2006}. This model is based on density-functional theory computed kinetic parameters of the set of 26 elementary processes defined by all non-correlated site and element specific adsorption, desorption, diffusion and reaction events that can occur on a lattice spanned by two different active sites offered by the surface. These two types of adsorption sites, the so-called bridge (br) and coordinatively unsaturated (cus) sites, are aligned in alternating rows on a simple square lattice, where the catalyst activity is most sensitive to the dynamics on the cus sites \cite{meskine2009}. The possible adsorbates are atomic oxygen on br and cus sites ($\rm O_{cus}$, $\rm O_{br}$) and CO on single br and cus sites ($\rm CO_{cus}$, $\rm CO_{br}$). For a given gas-phase impingement as characterized by the mass fractions, pressure and temperature at the surface, the 1p-kMC model yields the turnover frequencies averaged over mesoscopic areas as required for the boundary condition, Eq. (\ref{surfaceSource}) \cite{matera2010}. However, in contrast to prevalent microkinetic models based on mean-field rate equations these averages are properly derived by fully accounting for the microscopic site heterogeneities and chemical distributions at the surface \cite{temel2007,matera2011}.

Through the surface mass fractions in the TOFs the boundary condition, Eq. (\ref{surfaceSource}), actually depends on the very flow profile. In principle this dictates a simultaneous and self-consistent solution of flow equations and 1p-kMC, and this for every spatially resolved finite element cell at the surface. We decouple this otherwise numerically intractable problem through an instantaneous steady-state approximation \cite{deutschmann2008}, i.e. we assume that the surface chemistry adapts quasi-instantaneously on the time scales characteristic for any variations of the flow field. This allows to precompute the steady-state 1p-kMC TOFs for any reasonable impingement conditions, which then serve as a look-up table for the fluid dynamical simulations. In practice we generate this look-up table by employing a modified quadratic Shepard interpolation (MQSI) \cite{renka1988_1,renka1988_2} of the 1p-kMC raw data. Since re-adsorption of CO$_2$ is negligible at the RuO$_2$(110) surface \cite{reuter2006}, the TOF depends in the present problem only on temperature and the partial pressures of CO and O$_2$. For this three-dimensional (3D) interpolation we simply use gridded data as detailed before \cite{matera2010}. Nevertheless, MQSI is in principle a gridless scattered data interpolation technique. It should thus also work reasonably well in higher dimensions, i.e. for more complex reactions with more than two species, where grid-based data sets become impractical. MQSI interpolation and instantaneous steady-state decoupling provide thus a promising general route to integrate 1p-kMC simulations into fluid dynamical frameworks at computational costs comparable to conventional mean-field microkinetic formulations.

\subsection{Numerical solution}

Numerical solution of the balance equations, Eq. (\ref{Species}), is finally achieved with the adaptive finite element (FEM) code {\sc Kardos} \cite{kardos,lang2000}. For the channel we solve the 2D Cartesian version of Eq. (\ref{Species}) using dimensionless variables $x'=x/L$, $y'=y/L$, ${\bf v}'={\bf v}/u_{\rm max}$, and $t'=t u_{\rm max}/L$. For the stagnation flow we utilize the axial symmetry and solve the corresponding representation of Eq. (\ref{Species}) using the dimensionless variables $r'=r/L$, $y'=y/L$, ${\bf v}'={\bf v}/u_{\rm max}$ and $t'=t u_{\rm max}/L$. We employ (least-square) stabilized linear FEs in order to avoid the typical problems arising from the standard Galerkin method applied to convection dominated partial differential equations \cite{lang1998}. For the temporal discretization we use the {\sl ros1} time integrator, which was found to give the most stable numerical solution. The adaption of the grid in each time step is the most CPU-intensive part of the numerical solution. We therefore employ a two stage strategy: First the system is relaxed into steady state with only a very low numerical accuracy. Afterwards comparatively few further time steps are performed with successively increasing accuracy. After the desired accuracy is reached the grid is fixed (see the initial and the adapted grid for the channel flow shown in the lower two panel of Fig. \ref{fig1}). The problem adapted grid is then used to perform long time simulations to obtain accurate steady-state solutions with an estimated spatial accuracy of about $ 10^{-4}$. 

As for the ideal stagnation flow \cite{matera2010} both reactor-surface systems exhibit also a non-reactive steady state for a range of reaction conditions. In order to controllably reach the reactive steady state the temperature was linearly driven from $T=550$\,K to $T=600$\,K within 1\,s, employing time steps of $5 \times 10^{-5}$\,s for both cases. Afterwards the simulations are continued for another period of 24\,s, which is more than five times the apparent relaxation time to the steady state, i.e after roughly 4 seconds no further significant temporal variations have been observed. Since the gas phase above the catalyst is on average exchanged more than 100 times during this period we are confident that no oscillations with a longer period exist and the steady states discussed in the following are actually stable.

\section{Results}

\subsection{Considered reaction conditions}

The 1p-kMC model for the CO oxidation at RuO$_2$(110) and its results for the intrinsic activity have been discussed before \cite{reuter2006}. Under UHV operation conditions, where the intrinsic properties of the catalysts are not masked by macroscopic transport, the model reproduces experimental findings quantitatively \cite{rieger2008,reuter2004}. At ambient pressures the model produces three different surface phases: under O$_2$-rich feed the surface is fully O-covered by oxygen, while a CO-rich feed correspondingly leads to a CO-covered surface. Not surprising for a Langmuir-Hinshelwood type mechanism, highest turn-overs are observed for intermediate reaction conditions where both species are stabilized at the surface in appreciable amounts. It must be stressed that under CO-rich feeds the oxide surface could in reality be further reduced to a metal state. This phase transformation and any connected catalytic activity can not be treated by the present 1p-kMC model assuming an intact underlying RuO$_2$(110) lattice. Nevertheless the focus of this study is on the integration of a given 1p-kMC based microkinetic description into a computational treatment of macroscopic transport and the consequences resulting from the latter. For this purpose the employed model serves well enough, exhibiting all expected features of a high-TOF reaction like CO oxidation: (i) an intrinsic TOF narrowly peaked in $(T,p_{\rm O_2}, p_{\rm CO})$-space, and (ii) an insufficient description of this activity by standard rate equation based theories \cite{temel2007,matera2011}. If in future a refined microkinetic model will be developed this can be integrated in exactly the same way as the present one.

In both reactor geometries we illustrate the intricate coupling between surface chemistry and mass and momentum transfer by focusing the following nominal reaction conditions: At the inlet an essentially zero CO$_2$ concentration ($p^{\rm inlet}_{\rm CO_2} \equiv 10^{-5}$\,atm), partial pressures of O$_2$ and CO of $p^{\rm inlet}_{\rm O_2}=0.3$\,atm and $p^{\rm inlet}_{\rm CO}=1.8$\,atm, respectively, and a temperature of $T=600$\,K. These reaction conditions correspond to a low intrinsic activity with an almost fully O-covered surface, i.e. to a situation where one would intuitively not expect any mass transfer limitations. In the following we demonstrate that even there strong couplings arise, with the revealed effects likely to apply also to any other nominal reaction condition. 

\subsection{Concentration in the reactor}

\begin{figure}
\includegraphics[width=\linewidth]{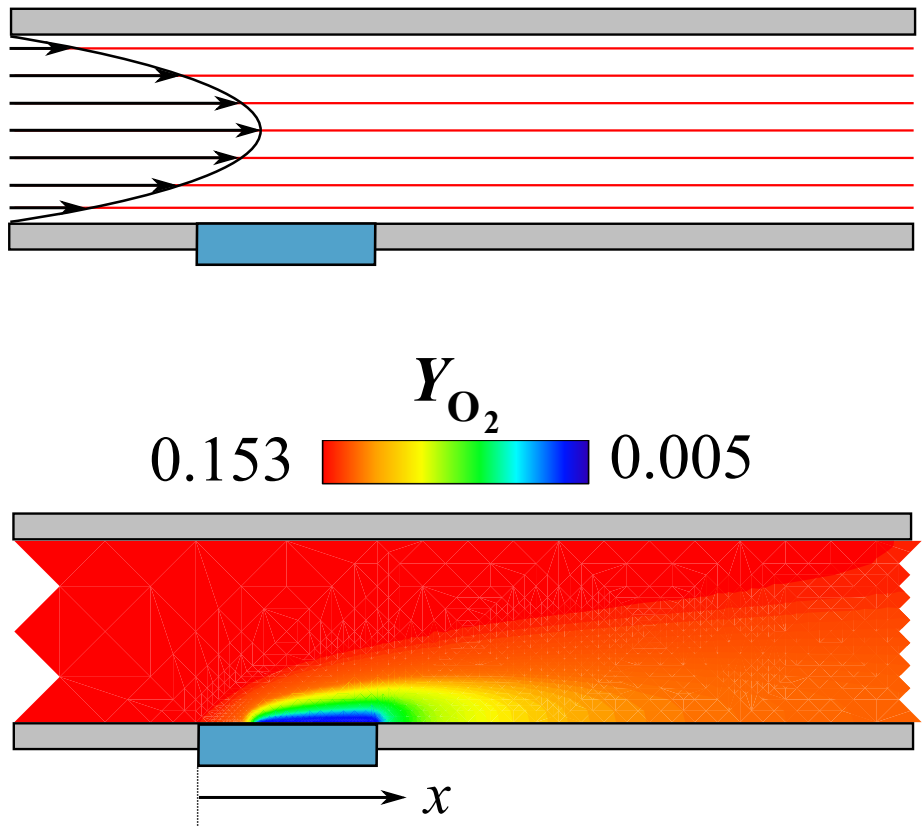}
\caption{(Color online) Channel flow: (Upper panel) Typical velocity profile (black) and stream lines (red); (lower panel) oxygen mass fraction. Strong variations by an order of magnitude are observed for the latter in the reactor and on the catalyst surface due to the feeding of the catalyst from the side.}
\label{fig3}
\end{figure}

\begin{figure}
\includegraphics[width=\linewidth]{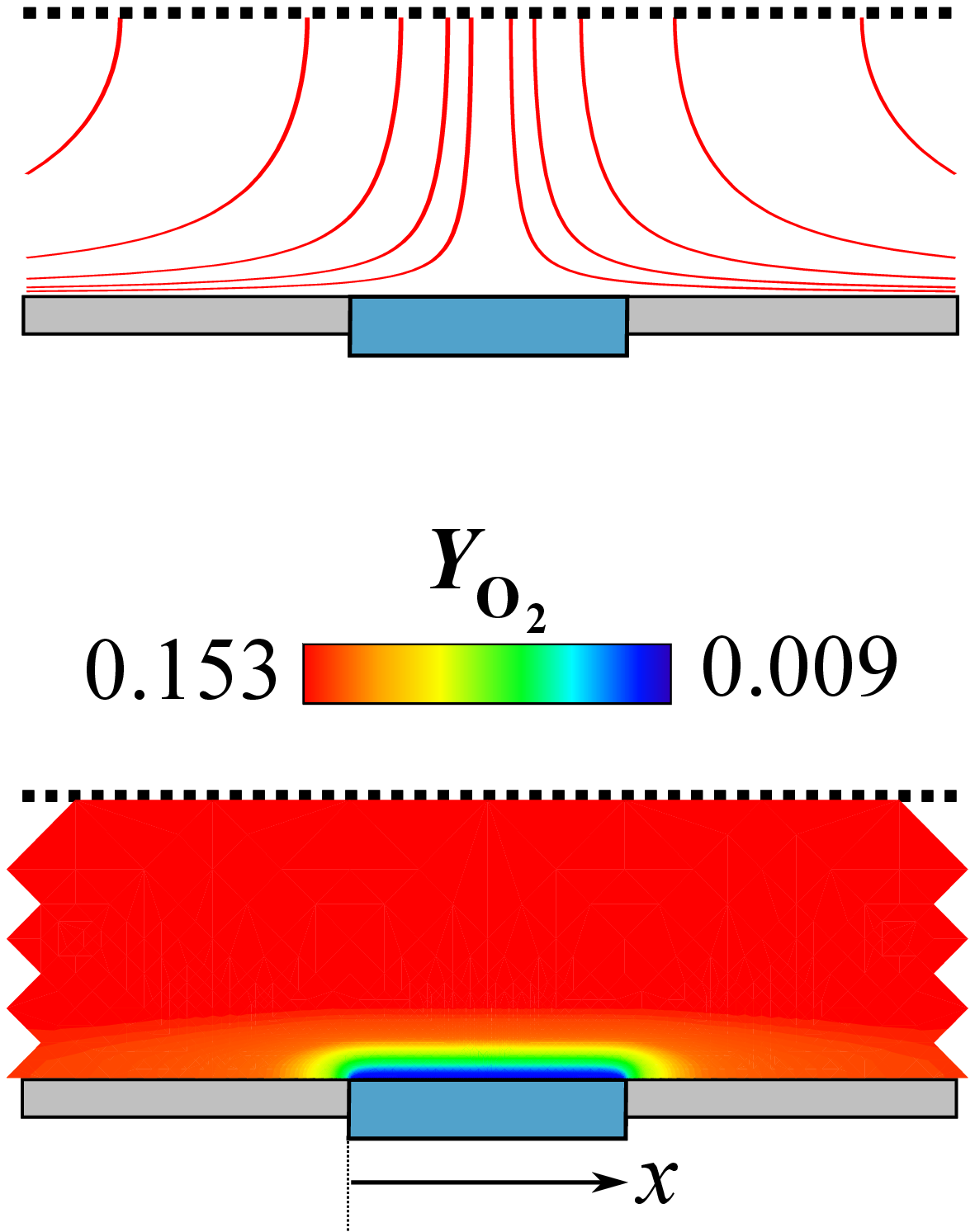}
\caption{(Color online) Stagnation flow: (Upper panel) Stream lines; (lower panel) oxygen mass fraction. Strong variations are observed for the latter in the reactor, but the catalyst surface experiences at least almost homogeneous reaction condition due to the feeding of the catalyst from the top.}
\label{fig4}
\end{figure}

For the chosen reaction conditions the resulting concentration profiles of O$_2$, i.e. the mass fraction of the minority species, in the two reaction chambers are shown in the lower panels in Fig. \ref{fig3} and Fig. \ref{fig4} for channel and stagnation flow, respectively. In both cases we show only the part of the reaction chamber with significant concentration variations. The upper panels in the corresponding figures show additionally typical streamline configurations for the respective flow geometry. The CO concentration behaves analogously to the oxygen concentration, but as it is the majority species in the chosen feed conditions the relative changes are much smaller. In both cases we find a significant deviation at the surface from the nominal applied reaction conditions at the inlet. The nominal oxygen mass fraction of 0.153 drops up to a factor of $\approx 25$ and $\approx 15$ for channel and stagnation flow, respectively. Excruciatingly, these large concentration variations of the order of a factor ten per millimeter would not be reflected by measurements taken at the outlets, as there the oxygen concentrations differ again only little from the nominal inlet values. In other words the observed mass transfer effects have a tremendous impact although we have only little conversion.

The origin of this unintuitive behavior is best explained for the channel flow. Entering at the inlet the gas maintains its nominal inlet composition until it actually reaches the catalyst. This originates from the dominant convective mass transport, pushing any concentration variation immediately downstream. In contrast, in the direct vicinity of the catalyst the ongoing surface chemistry induces a considerable lowering of the O$_2$ concentration and concomitantly also large concentration gradients, cf. Eq. (\ref{surfaceSource}). While the catalyst thus experiences a very oxygen-poor gas mixture the opposite reactor wall retains the nominal inlet composition, since the gas near this wall is transported through the area above the catalyst before it can reach the catalytic surface. Without a chemical source that induces concentration gradients, the mass fractions in the gas stream equilibrate downstream of the catalyst surface and thus the oxygen concentration increases again at the lower wall. 

The explanation of stagnation flow follows the same arguments, we just have curved stream lines, cf. upper panel of Fig. \ref{fig4}, instead of the straight stream lines from inlet to outlet as in the channel flow geometry (upper panel, Fig. \ref{fig3}). A stream line starting in the center of the inlet shows a similar behavior as a stream line in the lower part of the channel. Near the inlet convection dominates and we still observe the nominal composition. As the gas flows along the streamline it approaches the catalyst and concentration gradients are induced. A streamline further away from the center of the inlet never comes close to the catalyst surface and therefore behaves similar to a streamline in the upper part of the channel, passing the zone above the catalyst without being 
affected by the ongoing conversion. As in the channel flow concentration variations start to fade as soon as the gas leaves the area above the catalyst.

Despite the huge concentration variations within the reactor, the catalyst surface experiences a very homogeneous gas phase in the stagnation flow setup due to the broad front of reactants streaming perpendicularly against the catalyst and thereby continuously replenishing reactants at the nominal concentrations from the top. This is different for the channel flow. There the feeding of the catalyst from the side has the effect that reactants are constantly consumed while streaming along the catalyst surface, and, therefore, the largest spatial variations occur actually on the catalyst surface.

\subsection{At the surface}

\begin{figure}
\includegraphics[width=\linewidth]{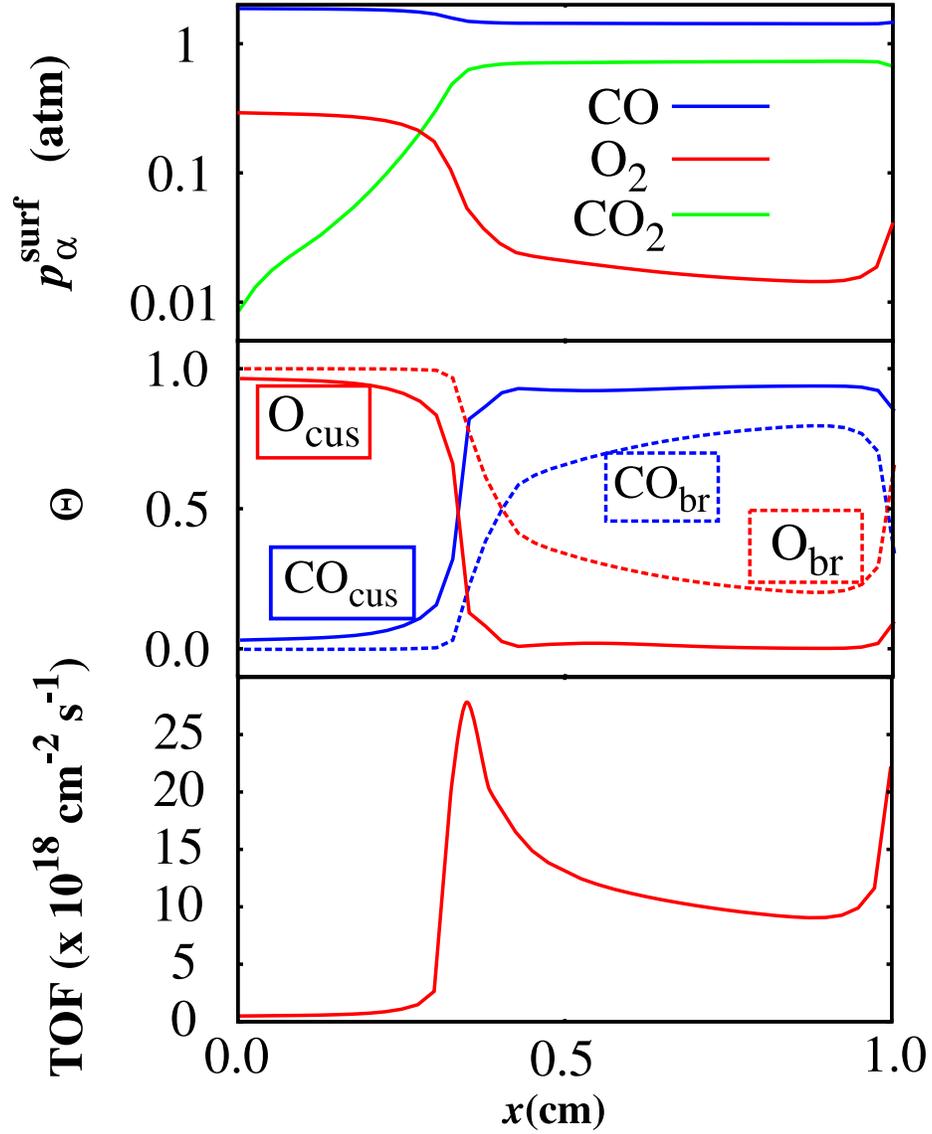}
\caption{(Color online) Channel flow: Partial pressures (upper panel), coverages (middle panel) and local TOF (lower panel) directly at the catalyst surface and spatially resolved across the lateral width $L=1.0$\,cm of the model catalyst, cf. Fig. \ref{fig1}. All displayed quantities show tremendous variations along the catalyst surface, cf. Fig. \ref{fig3} for the definition of the lateral coordinate $x$ employed.}
\label{fig5}
\end{figure}

\begin{figure}
\includegraphics[width=\linewidth]{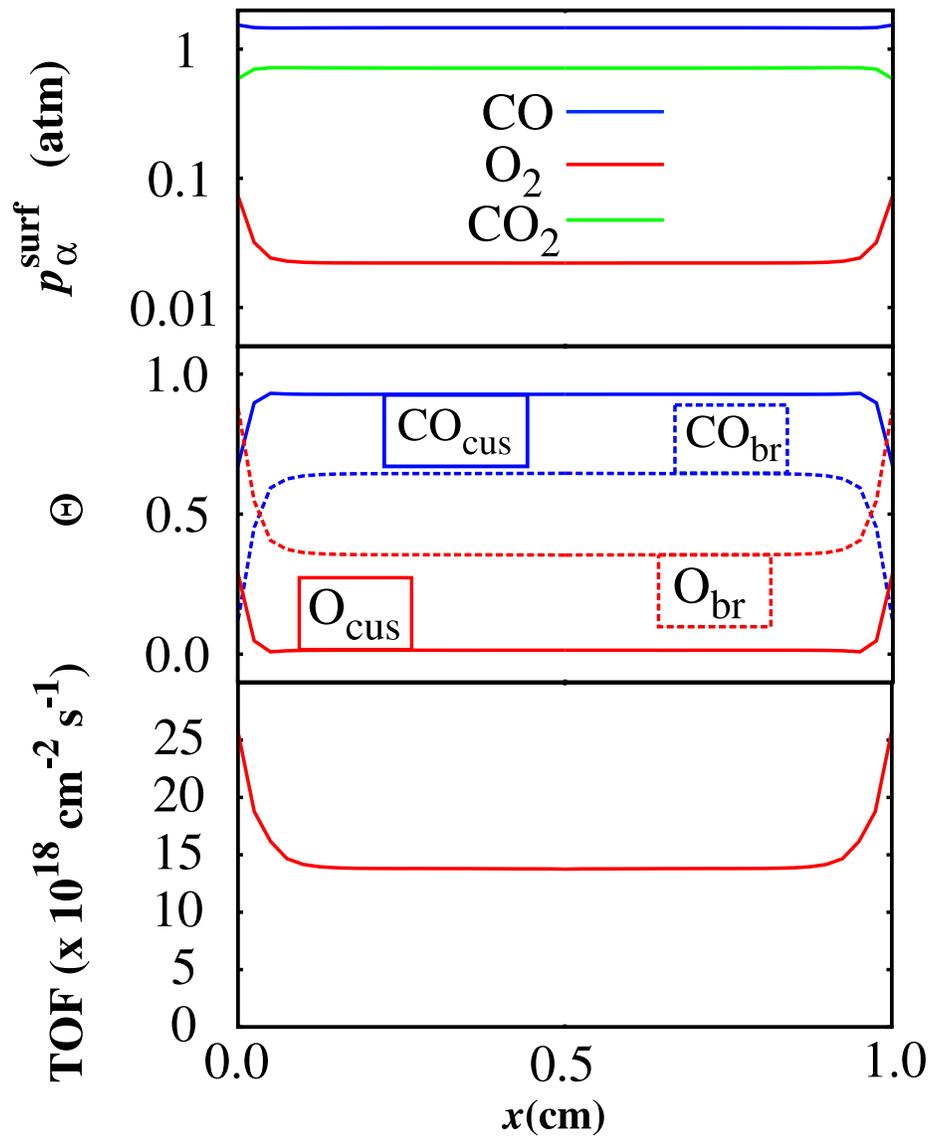}
\caption{(Color online) Same as Fig. \ref{fig5}, but for the stagnation flow. Different from the channel flow there are now no lateral variations in the central region of the catalyst, i.e. variations are only observed at the edges of the catalyst disc, cf. Fig. \ref{fig4} for the definition of the lateral coordinate $x$ employed.}
\label{fig6}
\end{figure}

So far, we have discussed the influence of the catalyst activity on the flow profiles. With the target of obtaining atomistic insight into the operating catalyst, the central interest in {\em in-situ} experiments is instead the reverse, i.e. what is actually going on at the surface. In Figs. \ref{fig5} (channel flow) and \ref{fig6} (stagnation flow) we therefore display the dependence on the position on the catalyst surface of the partial pressures $p^{\rm surf}_\alpha$ ($\alpha \in \{{\rm CO, O_2, CO_2}\}$), coverages $\Theta_{\beta}$ at the two different adsorption sites ($\beta \in \{{\rm O_{cus}, O_{br}, CO_{cus}, CO_{br}}\}$), and the TOF. The lateral variable $x$ employed runs hereby from the left to right edge of the catalyst as sketched in Figs. \ref{fig3} and Fig. \ref{fig4}, respectively.

In the channel we have relatively high partial pressures of CO and oxygen at the upstream (left) edge of the catalyst. Following the stream from left to right both partial pressures first slowly decrease, with a concomitant increase of the CO$_2$ partial pressure. As soon they reach a certain critical value, both partial pressures drop down sharply. After that oxygen has almost vanished, while the majority CO species plateaus at a relatively high value, corresponding to the unconsumed fraction. Further downstream only little conversion occurs, until at the downstream edge the partial pressures of the reactants increase again.  
This behavior can be nicely rationalized in terms of the coupling of macroscopic mass transport and mesoscopic surface chemistry. For the nominal reaction 
conditions the 1p-kMC model predicts an almost fully oxygen-covered surface and corresponding low reactivity. At the catalyst left edge ($x=0$) this nominal behavior is still observed. Due to the low reactivity the partial pressures of CO and O$_2$ initially decrease only slowly while the gas streams along the surface. However, since oxygen is the minority species, its relative change is larger than that of CO. As even at this low reactivity, the gas transport can not maintain the nominal mixture, the catalyst experiences an increasingly CO-rich gas phase. The concomitant increase of CO coverage at the surface goes hand in hand with an increase of the local TOF, as expected for a Langmuir-Hinshelwood type mechanism. At $x \approx 0.3$\,cm the surface ultimately changes from the O-poisoned to a well mixed state. This well-mixed state exhibits a very high intrinsic activity, and correspondingly a steep increase in the local TOF results. Due to this high reactivity, oxygen is quickly consumed in the gas-phase above the catalyst, and under the present mass transfer limitations the oxygen partial pressure drops steeply. Just 1\,mm further downstream of this turning point, the gas stream is then largely CO-dominated and thus we observe an almost zero coverage of oxygen on cus-sites. Since the formation of CO$_2$ on RuO$_2$ is dominated by reaction paths involving the O$_{\rm cus}$ species, the reactivity  falls off again \cite{meskine2009}. This decrease in oxygen partial pressure, reactivity and oxygen coverage continues until $x\approx 0.9$\,cm, where the oxygen partial pressures reaches its minimum of $1.5\times 10^{-2}$\,atm. However, the reactivity does not drop down to the very low TOFs observed at the catalyst left edge for the O-covered surface. This is inherent to the intrinsic reactivity of the RuO$_2$ microkinetic model, as surface diffusion limitations lead to particularly low TOFs for the O-poisoned state \cite{matera2011}. Last, the increase of reactant concentration for $x > 0.9$\,cm is due to the equilibration of concentration variations behind the catalyst and the concomitant increase of reactant concentration at the lower reactor wall. Since the velocity approaches zero at this wall the resulting diffusive mass flux pointing towards the low concentration region can effectively transport reactants upstream. As O$_2$ is the minority species, the relative increase of its concentration is again larger than for CO, leading to an increase of the O$_{\rm cus}$ coverage together with an increase of the reactivity as before.

In contrast to the channel flow, the stagnation flow exhibits only little lateral variations of the quantities displayed in Fig. \ref{fig6} and only close to the catalyst edges, at $x=0.0$ and $x=1.0$\,cm. Notwithstanding, due to mass transfer limitations the gas-phase composition still deviates strongly from the nominal one at the inlet. In other words, the whole center part of the catalyst between $x \approx 0.2$\,cm and $x \approx 0.8$\,cm experiences roughly the same gas phase, but this gas phase is different to the one at the inlet. Again, the changes in the partial pressure are much larger for the O$_2$ minority species, with the $O_2$ partial pressure going down to $2.2\times10^{-2}$\,atm in the center of the disc.
In the resulting very CO-rich composition, the surface is mostly covered with CO and the TOF is medium high as in the downstream half of the catalyst in the channel flow geometry. Similarly to the channel flow, we observe an increase of the reactant partial pressures at the catalyst edges due to upstream diffusion from the area outside the region above the catalyst, where concentration gradients equilibrate and the reactant concentration is higher than at the surface. Since this has stronger relative effect on the oxygen the surface is covered with more oxygen, especially at the cus sites, and the local reactivity increases towards the downstream edges, just as in the channel flow case.

\section{Discussion}

\begin{figure}
\includegraphics[width=\linewidth]{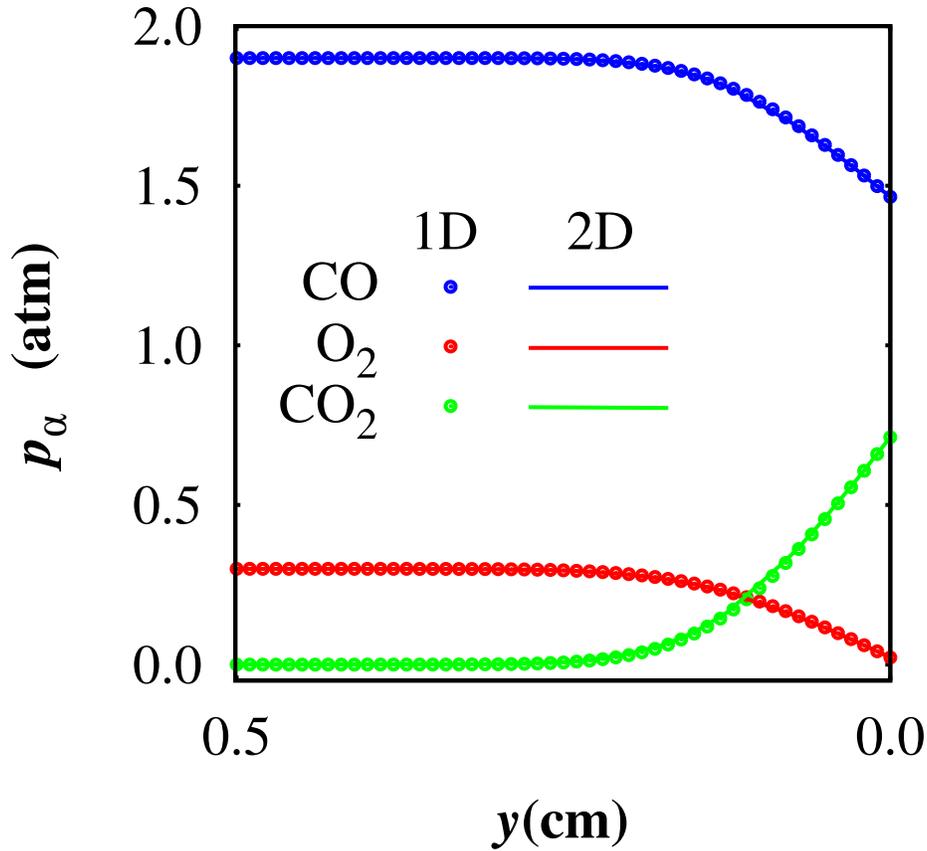}
\caption{(Color online) Partial pressure profiles along the symmetry axis, cf. Fig. \ref{fig2}, as obtained from the 1D ideal stagnation flow (points) and the full axisymmetric problem (lines), see text. $y=0.0$ corresponds to the catalyst surface and only the 5\,mm above the catalyst are shown, as significant pressure variations are restricted to this region.
}\label{fig7}
\end{figure}

The presented results highlight the difficulties when performing {\em in-situ} atomic scale resolution experiments in badly controllable flow geometries like the channel flow. For the latter geometry, the dramatically varying lateral surface concentrations shown in Fig. \ref{fig3} demonstrate that local atomic-scale resolving experiments can arrive at completely different, if not contradictory results, depending on where they are positioned over the catalyst. Furthermore, reaction conditions and reactivity might vary by orders of magnitude along the catalyst surface. As a consequence the targeted correlation of atomic-scale structure, reactivity and reaction conditions might become impossible. Even worse, not accounting for the possible interaction between transport and chemistry can easily lead to completely wrong conclusions. For instance, a local measurement at $x=0.2$\,cm in the presented channel flow problem will yield an O-covered surface. Without considering transport and transport-induced variations across the catalyst surface, a high average turnover will be deduced from the measured conversions at the outlet. This will then be assigned to this surface state, which is in reality hardly active. In the worst case, this assignment of a high activity to an O-poisoned state might spur conjectures of an active Eley-Rideal mechanism, instead of the true Langmuir-Hinshelwood mechanism underlying the employed microkinetic model. 

The surface science field is full with such conjectures, and in this respect our results provide a new perspective on the discussions on the active state of transition metal surfaces in CO oxidation catalysis, here particularly ruthenium (see e.g. \cite{Goodman2007,Over2007,Goodman2007_2}). In the downstream half of the discussed channel flow problem, we have an excess of gas-phase CO and a CO-covered surface. Under these conditions the catalyst will in reality be reduced to the metallic state. As mentioned above, we cannot account for this within the employed microkinetic model, however, for the transport-limitation argument this makes little difference. An atomic-scale investigation e.g. with reactor-STM or SXRD in this region would suggest to assign observed (average) reactivity to the reduced catalyst, not aware that the observed reactivity might in reality originate from the upstream parts, which are less likely to be reduced. In principle, one might equally conceive the opposite case, an active pristine metal in a gas stream with oxygen excess. In this case, CO would be the minority species with
transport limitations possibly inducing a huge drop in its partial pressure over the catalyst surface. In the increasingly O-rich environment the surface would be oxidized in the downstream part of the catalyst. With the same local measurement the reactivity would then erroneously be assigned to the oxide, even so it might in reality be hardly active under oxygen excess conditions.

As demonstrated in the previous section lateral macroscopic heterogeneity is not such a big problem in the case of stagnation flows, if one concentrates on the central part of the catalyst. For the discussed reaction conditions, average TOFs monitored at the outlet would also agree much better with the local TOF in the center of the catalyst, as the TOF increase at the edges is only moderate and will not falsify the average TOF much. In general, this must not necessarily be so though, i.e. the average TOF might be dominated by edge effects. In this situation, it can be preferable to rather extract the local TOF in the central region directly from measured concentration profiles, i.e. without necessity of an averaging procedure. For the stagnation flow such a procedure is possible due to the homogeneity in the central part, i.e. above the center of the catalyst there are only variations in axial ($y$) direction, cf. Fig. \ref{fig2}. Under these conditions the coupled flow problem, i.e. the full cross-coupled Navier-Stokes and species transport equation Eq. (\ref{Species}), can be reduced to the 1D boundary value problem describing the ideal stagnation flow \cite{kee2003}, which we have previously discussed for the present CO oxidation at RuO$_2$(110) \cite{matera2009,matera2010}. In Fig. \ref{fig7} we compare the results from these ideal stagnation flow equations with the 2D model employed here. Not only for the displayed reaction conditions, the resulting partial pressure profiles along the symmetry axis $y$ in the boundary layer above the catalyst show very good agreement. To one end this validates that the area above the central part of the catalyst can indeed be modeled to high accuracy with the simple 1D stagnation flow equations. Since these stagnation flow equations explicitly include the coupling between chemistry and Navier-Stokes equation, these findings furthermore justify the present neglect of this coupling, i.e. the use of a predefined velocity field. With the equivalence of both approaches for the central part of the catalyst established, we realize that one local TOF value is sufficient to close the idealized 1D flow equations. Rather than providing this one value (in the present case through the microkinetic model), a local gas-phase composition directly at the surface might equally be employed. If this quantity is measured experimentally, e.g. through local mass spectroscopy \cite{mcguire2011}, laser-induced fluorescence \cite{gudmundson1998}, or simply by drilling a hole into the center of the catalyst sample and extracting the gas for a composition analysis there, then the corresponding local TOF can be determined through the solution of the 1D stagnation flow equations and exploiting Eq. (\ref{surfaceSource}).

Such an approach is unfortunately only possible for the axisymmetric stagnation flow. In less well controlled geometries like the channel flow the need to determine a 1D or 2D function as boundary condition requires an elegant parametrization of this function, today almost exclusively in terms of a mean-field based (micro-)kinetic model. The resulting fitting problem is computationally much more challenging due to the increase in parameter space and second due to the by far more costly multi-dimensional simulations. Further it is also error-prone, since it relies on the accuracy of the employed kinetic model. The stagnation flow instead allows to obtain the necessary kinetic data without any assumptions of the mechanism or even the structure of the catalyst. We acknowledge that the stagnation flow might not be suitable for particular {\em in-situ} experiments, e.g. due to the need of differential pumping in photo-electron spectroscopies
or the need of a tip in surface scanning experiments. Nevertheless, the obtained kinetic data from a stagnation flow reactor can be used to determine the local reaction conditions and reactivity in other flow geometries numerically, for instance by interpolating the data in a similar fashion as we have done for the 1p-kMC data. Doing so one is able to correlate reactivity and partial pressures at the surface (from simulation) with the atomic scale insight from experiment, thereby allowing a by far more qualified discussion and making the findings from different {\em in-situ} experiments much better comparable.

\section{Conclusion}

We have extended our multi-scale methodology initially developed for a quasi 1-dimensional stagnation flow problem to 2D flow geometries. Within the employed instantaneous steady-state approximation the additional complexity of more general flows solely manifests itself in the numerical solution of the gas flow, not
in the necessity to model more surface points with 1p-kMC microkinetic simulations. The presented approach can therefore readily be applied to even more
complex geometries, and is in principle only limited by the capabilities of the used flow solvers. Due to its modular structure it can also easily be coupled
to existing Computational Fluid Dynamics packages with only a minimal effort in programming and code development.

Specifically we have compared a two-dimensional channel flow with an axisymmetric stagnation flow geometry for the CO oxidation at a RuO$_2$(110) model catalyst surface, as rather simple albeit representative showcase for multi-dimensionality of the concentration profiles occurring in {\em in-situ} reaction chambers. Similar to the ideal quasi-one-dimensional stagnation flows \cite{matera2009,matera2010} we find large deviations between the nominal applied reactions conditions and those experienced by the catalyst surface. 

Moreover, the simulations reveal that large lateral variations of the concentrations at the catalyst surface can arise in the channel flow geometry. These can then be accompanied by corresponding changes in the surface coverages and the local TOF. While in the channel flow these variations extend over the whole surface, they are limited to the part near the catalyst's edge in the case of the stagnation flow. This behavior allows for the application of an effective 1D model, and thereby opens the way to a model-free determination of kinetic data. 

In general, the results presented clearly demonstrate the need for an integrated multi-scale modeling of surface chemistry and fluid flow, when trying to interpret today's {\em in situ} experiments on the basis of microscopic simulations. On the experimental side they show the danger of misinterpreting the results from {\em in situ} experiments. Depending on which part of the catalyst is investigated by an atomic-resolution technique like reactor STM or SXRD very different microscopic states can be observed. Clearly there is the need for well controlled flow geometries and/or spatially resolved measurements of gas phase concentration to further advance this field.

\end{document}